\def\ni{\noindent}
\def\vs{\vskip.3cm}

\def\+{{(+)}}  \def\-{ {(-)} }   \def\0{ {(0)} }
\def\1{ {(1)} }  \def\2{ {(2)} }

\def\sq{Q\kern-6pt/}
\def\sQ{Q\kern-12pt\nearrow}
\documentclass[11pt,eqs]{article}

\usepackage{latexsym}
\usepackage{amssymb}

\def\be{\begin{equation}}             \def\ee{\end{equation}}
\def\ba{\begin{array}{rcl}}           \def\ea{\end{array}}
\def\beqa{\begin{eqnarray} }          \def\eeqa{\end{eqnarray} }
\def\beqalign{\begin{eqalign}}        \def\eeqalign{\end{eqalign}}
             
\def\bsubeq{\begin{subequations}}     \def\esubeq{\end{subequations}}
\def\bitem{\begin{itemize}}           \def\eitem{\end{itemize}}

\def\DJ{\leavevmode\setbox0=\hbox{D}\kern0pt
 \rlap{\kern.04em\raise.188\ht0\hbox{-}}D}
\def\dj{\leavevmode\setbox0=\hbox{d}\kern0pt
 \rlap{\kern.215em\raise.46\ht0\hbox{-}}d}

\newcommand{\bd}{\begin{displaymath}}
\newcommand{\ed}{\end{displaymath}}

\begin{document}

\title{ Fermionic T-duality and momenta noncommutativity
\thanks{Work supported in part by the Serbian Ministry of Education and Science, under contract No. 171031.}}
\author{B. Nikoli\'c \thanks{e-mail address: bnikolic@ipb.ac.rs} and B. Sazdovi\'c
\thanks{e-mail address: sazdovic@ipb.ac.rs}\\
       {\it Institute of Physics}\\{\it University of Belgrade}\\{\it P.O.Box 57, 11001 Belgrade, Serbia}}
\maketitle

\begin{abstract}

In this article we establish the relationship between
fermionic T-duality and momenta noncommuativity. This is extension of known relation between bosonic T-duality and coordinate noncommutativity. The case of open string propagating in background of the type IIB superstring theory has been considered. We perform T-duality with respect to the fermionic variables instead to the bosonic ones. We also choose
Dirichlet boundary conditions at the string endpoints, which lead to the momenta noncommutativity, instead Neumann ones which lead to the coordinates noncommutativity. Finally, we establish the main result of the article that
momenta noncommutativity parameters are just fermionic T-dual fields.

\end{abstract}
\vs

\ni {\it PACS number(s)\/}: 11.10.Nx, 04.20.Fy, 11.10.Ef, 11.25.-w   \par

\section{Introduction}
\setcounter{equation}{0}

Two theories that are
dual to one another can be viewed as being physically identical  \cite{jopol}. An important
kind of duality is so called T-duality, where T stands for
target space-time. This means that we
can switch the target space with its dual without loosing the physical content
of the theory.

When the open string endpoints are attached to D-brane, its world-volume becomes noncommutative manifold \cite{5,6,7,kanonski, BNBS}. The noncommutativity parameter is proportional to the Neveu-Schwarz antisymmetric field $B_{\mu\nu}$, while in the supersymmetric case the noncommutative parameters are proportional to the $\Omega$ odd parts of NS-R field, $\Psi^\alpha_{-\mu}$, and R-R field strength, $F_{s}^{\alpha\beta}$.  The noncommutative (super)space is startting point in studying properties of the noncommutative (super) Yang-Mills theories \cite{voja}.

Recently a new kind of T-duality was discovered, the fermionic
T-duality \cite{ferdual}. It consists in certain non-local redefinitions of the fermionic variables of the superstring mapping a supersymmetric background to another supersymmetric background. Technically fermionic T-duality is
similar to the bosonic one, except that dualization is performed
along fermionic directions, $\theta^\alpha$ and
$\bar\theta^\alpha$. Ref.\cite{ferdual} also shows that T-duality maps gluon scattering amplitudes  in the original theory to Wilson loops in the dual theory. They also investigated connection between "dual conformal symmetry" and integrability. The articles {\cite{MW1}}, focussing more on integrability, deal with fermionic T-duality also, using Green-Schwarz string on $AdS_5\times S^5$. From slightly different point of view most of the results of the Ref.\cite{ferdual} have been obtained.

The present article is motivated by the fact that for the specific
solution of the boundary conditions some of the bosonic T-dual
background fields coincide with noncommutativity
parameters \cite{bnbsjhep, bnbsnpb}. In these articles type IIB
superstring theory in pure spinor formulation has been considered.
Performing Buscher T-duality \cite{buscher} along all bosonic
directions $x^\mu$, the background fields of the T-dual theory
have been found. On the other hand, consequences of the particular
boundary conditions at the open string endpoints have been
investigated: the Neumann boundary conditions for bosonic
coordinates and preserving half of the initial $N=2$ supersymmetry
for fermionic ones. It turned out that coordinates
noncommutativity parameters are the bosonic T-dual fields. So, the
particular choice of duality (along bosonic directions)
corresponds to the particular choice of boundary conditions. In the present article we are looking for
such boundary conditions which produce noncommutativity parameters
equal to the fermionic T-dual background fields.

The article is organized in the following way. First, we introduce
the action of the pure spinor formulation for type IIB superstring
theory keeping quadratic terms. Then, we perform canonical
analysis in the light-cone coordinates. Because of
reparameterization invariance we can take any timelike or
lightlike coordinate as evolution parameter. For lightlike
evolution parameter the Lagrangian is linear in velocities, and
there are primary constraints which we will use as suitable
introduced currents. There are two cases for consideration: 1)
$\tau\to\sigma_-$ and $\sigma\to\sigma_+$ and 2) $\tau\to\sigma_+$
and $\sigma\to-\sigma_-$. Canonical Hamiltonian with timelike
evolution parameter can be written in the Sugawara form of the
currents.

In the case of open string action principle, besides equations of motion, produces boundary conditions. Choosing Dirichlet boundary conditions and
treating them as canonical constraints \cite{kanonski,BNBS}, we obtain the initial
coordinates and momenta in terms of the effective ones, which are odd under world-sheet parity transformation $\Omega:\sigma\to-\sigma$. It turns out that momenta
are noncommutative, while the coordinates are commutative. The source of noncommutativity is the presence of the effective coordinates
in the solution for initial momenta. The noncommutativity
parameters are fermionic T-dual background fields.

At the end we give some concluding remarks.

\section{Type IIB superstring and fermionic T-duality}
\setcounter{equation}{0}

In this section we will introduce the action of type IIB
superstring theory in pure spinor formulation and perform
fermionic T-duality \cite{ferdual}.

The action of type IIB superstring theory in pure
spinor formulation (up to the quadratic terms \cite{berko,susyNC,BNBSPLB, bnbsjhep, bnbsnpb} and neglecting ghost
terms as in Ref.\cite{susyNC}) is of the form
\begin{eqnarray}\label{eq:SB}
&{}&S=\kappa \int_\Sigma d^2\xi \partial_{+}x^\mu
\Pi_{+\mu\nu}\partial_- x^\nu \\&+&\int_\Sigma d^2 \xi \left[
-\pi_\alpha
\partial_-(\theta^\alpha+\Psi^\alpha_\mu
x^\mu)+\partial_+(\bar\theta^{\alpha}+\bar \Psi^{\alpha}_\mu
x^\mu)\bar\pi_{\alpha}+\frac{1}{2\kappa}\pi_\alpha F^{\alpha
\beta}\bar \pi_{\beta}\right ]\, ,\nonumber
\end{eqnarray}
where the world sheet $\Sigma$ is parameterized by
$\xi^m=(\xi^0=\tau\, ,\xi^1=\sigma)$ and
$\partial_\pm=\partial_\tau\pm\partial_\sigma$. Superspace is
spanned by bosonic coordinates $x^\mu$ ($\mu=0,1,2,\dots,9$) and
fermionic ones, $\theta^\alpha$ and $\bar\theta^{\alpha}$
$(\alpha=1,2,\dots,16)$. The variables $\pi_\alpha$ and $\bar
\pi_{\alpha}$ are canonically conjugated momenta to
$\theta^\alpha$ and $\bar\theta^\alpha$, respectively. All spinors
are Majorana-Weyl ones and $\Pi_{\pm
\mu\nu}=B_{\mu\nu}\pm\frac{1}{2}G_{\mu\nu}$.

On the equations of motion for fermionic momenta $\pi_\alpha$ and $\bar\pi_\alpha$ we obtain
\begin{equation}\label{eq:impulsi}
\pi_\alpha=-\frac{1}{2}\partial_+\bar\eta_\alpha\, ,\quad \bar\pi_\alpha=\frac{1}{2}\partial_-\eta_\alpha\, ,
\end{equation}
where we introduce useful notation
\begin{equation}\label{eq:etas}
\eta_\alpha\equiv4\kappa (F^{-1})_{\alpha\beta}(\theta^\beta+\Psi^\beta_\mu x^\mu)\, ,\quad \bar\eta_\alpha\equiv 4\kappa (\bar\theta^\beta+\bar\Psi^\beta_\mu x^\mu)(F^{-1})_{\beta\alpha}\, .
\end{equation}
Using these relations the action gets the form
\begin{eqnarray}\label{eq:lcdejstvo}
S(\partial_\pm x, \partial_- \theta, \partial_+
\bar\theta)&=&\kappa \int_\Sigma d^2\xi \partial_+ x^\mu
\Pi_{+\mu\nu}\partial_-x^\nu\\&+&2 \kappa \int_\Sigma d^2\xi \partial_+\left(\bar\theta^\alpha+\bar\Psi^\alpha_\mu x^\mu\right)
(F^{-1})_{\alpha\beta}\partial_-\left(\theta^\beta+\Psi^\beta_\nu x^\nu\right)\, .\nonumber
\end{eqnarray}

Now we will perform fermionic T-duality presented in
Ref.\cite{ferdual}. We suppose that the action has a global shift
symmetry in $\theta^\alpha$ and $\bar\theta^\alpha$ directions.
So, we introduce gauge fields $(v^\alpha_+, v^\alpha_-)$ and
$(\bar v^\alpha_+, \bar v^\alpha_-)$ and make a change in the action
\begin{equation}
\partial_-\theta^\alpha \to D_- \theta^\alpha\equiv\partial_-\theta^\alpha+v_-^\alpha\, , \quad \partial_+\bar\theta^\alpha \to D_+
\bar\theta^\alpha\equiv\partial_+\bar\theta^\alpha+\bar v_+^\alpha\, .
\end{equation}
In addition we introduce the Lagrange multipliers
$\vartheta_\alpha$ and $\bar\vartheta_\alpha$ which will impose that
field strengths of gauge fields $v_\pm^\alpha$ and $\bar
v_\pm^\alpha$ vanish
\begin{equation}
S_{gauge}(\vartheta,v_\pm,\bar \vartheta,\bar v_\pm)=\frac{1}{2}\kappa \int_\Sigma d^2\xi \bar \vartheta_\alpha (\partial_+
v_-^\alpha-\partial_- v^\alpha_+)+\frac{1}{2}\kappa \int_\Sigma d^2\xi  (\partial_+
\bar v_-^\alpha-\partial_- \bar v^\alpha_+)\vartheta_\alpha\, ,
\end{equation}
and the full action is of the form
\begin{equation}\label{eq:auxdejstvo}
S^\star(x, \theta,\bar\theta,\vartheta,\bar \vartheta,v_\pm,\bar v_\pm)=S(\partial_\pm x, D_- \theta, D_+ \bar\theta)+S_{gauge}(\vartheta,\bar \vartheta, v_\pm, \bar v_\pm)\, .
\end{equation}
If we vary with respect to
the Lagrange multipliers $\vartheta_\alpha$ and $\bar \vartheta_\alpha$ we obtain $\partial_+
v^\alpha_--\partial_- v^\alpha_+=0$ and $\partial_+
\bar v^\alpha_--\partial_- \bar v^\alpha_+=0$ which gives
\begin{equation}\label{eq:vteta}
\bar v_\pm^\alpha=\partial_\pm \bar\theta^\alpha\, ,\quad v_\pm^\alpha=\partial_\pm \theta^\alpha\, .
\end{equation}
Substituting these expression in (\ref{eq:auxdejstvo}) we obtain the initial action (\ref{eq:lcdejstvo}).

Now we can fix $\theta^\alpha$ and $\bar\theta^\alpha$ to zero and obtain the action quadratic in
the fields $v_\pm$ and $\bar v_\pm$
\begin{eqnarray}
&{}&S^\star=\kappa \int_\Sigma d^2\xi \partial_+ x^\mu \left[\Pi_{+\mu\nu}+2\bar\Psi^\alpha_\mu(F^{-1})_{\alpha\beta}\Psi^\beta_\nu\right]\partial_-x^\nu \\ &{}& +2\kappa \int_\Sigma \left[ \bar v_+^\alpha (F^{-1})_{\alpha\beta}v_-^\beta+\bar v_+^\alpha (F^{-1})_{\alpha\beta}\Psi^\beta_\nu\partial_-x^\nu+\partial_+x^\mu \bar\Psi^\alpha_\mu (F^{-1})_{\alpha\beta}v_-^\beta\right]\nonumber \\ &{}& +\frac{1}{2}\kappa \int_\Sigma d^2\xi \bar \vartheta_\alpha (\partial_+
v_-^\alpha-\partial_- v^\alpha_+)+\frac{1}{2}\kappa \int_\Sigma d^2\xi  (\partial_+
\bar v_-^\alpha-\partial_- \bar v^\alpha_+)\vartheta_\alpha\, ,\nonumber
\end{eqnarray}
which can be integrated out classically.
On the equations of motion for $v_\pm$ and $\bar v_\pm$ we obtain, respectively
\begin{equation}\label{eq:jed1}
\partial_- \bar \vartheta_\alpha=0\, ,\quad \bar v_+^\alpha=\frac{1}{4}\partial_+ \bar \vartheta_\beta F^{\beta\alpha}-\partial_+ x^\mu \bar\Psi^\alpha_\mu\, ,
\end{equation}
\begin{equation}\label{eq:jed2}
\partial_+ \vartheta_\alpha=0\, ,\quad v_-^\alpha=-\frac{1}{4}F^{\alpha\beta}\partial_-\vartheta_\beta-\Psi^\alpha_\mu \partial_- x^\mu\, .
\end{equation}

Substituting these expression in the action $S^{\star}$
we obtain the dual action
\begin{eqnarray}
&{}& {}^\star S(\partial_\pm x, \partial_- \vartheta, \partial_+
\bar \vartheta)=\kappa\int_\Sigma d^2\xi \partial_+ x^\mu  \Pi_{+\mu\nu}\partial_- x^\nu\, ,\\ &{}&+\frac{\kappa}{8}\int_\Sigma d^2\xi\left[\partial_+\bar \vartheta_\alpha F^{\alpha\beta}\partial_-\vartheta_\beta -4\partial_+x^\mu\bar\Psi^{\alpha}_{\mu} \partial_-\vartheta_\alpha+4\partial_+\bar \vartheta_\alpha\Psi^{\alpha}_{\mu}\partial_- x^\mu\right]\, ,\nonumber
\end{eqnarray}
from which we read the dual
background fields (denoted by stars)
\begin{equation}\label{eq:GBdual}
{}^\star B_{\mu\nu}=B_{\mu\nu}+\left[ (\bar\Psi F^{-1}\Psi)_{\mu\nu}-(\bar\Psi F^{-1}\Psi)_{\nu\mu}\right] \, , {}^\star G_{\mu\nu}=G_{\mu\nu}+2\left[ (\bar\Psi F^{-1}\Psi)_{\mu\nu}+(\bar\Psi F^{-1}\Psi)_{\nu\mu}\right]\, ,
\end{equation}
\begin{equation}\label{eq:Psidual}
{}^\star\Psi_{\alpha \mu}=4(F^{-1}\Psi)_{\alpha\mu}\, ,\quad {}^\star\bar\Psi_{\mu\alpha}=-4(\bar\Psi F^{-1})_{\mu\alpha}\, ,
\end{equation}
\begin{equation}\label{eq:Fdual}
{}^\star F_{\alpha\beta}=16(F^{-1})_{\alpha\beta}\, .
\end{equation}
Let us note that two successive dualizations give the initial
background fields.

\section{Canonical structure of the theory}
\setcounter{equation}{0}

The main technical problem is to perform complet consistency procedure
for the constraints because it has infinite many steps. So, it is useful to find
such basic variables (currents), which Poisson brackets with
Hamiltonian are as simple as possible. Following the idea of
Ref.\cite{BPS}, we can obtain these currents as canonical
constraints when lightlike direction is evolution parameter. It
turns that they are good basis for all canonical supervariables,
and that they have simple Poisson brackets as well with Hamiltonian
as among them.

\subsection{Canonical analysis with light-like evolution parameter}

Because of world sheet reparametrization invariance, any timelike or lightlike coordinate could be chosen as evolution parameter. The action
(\ref{eq:lcdejstvo}) is linear in derivatives with respect to the
light-cone coordinates $\partial_\pm$. So, in order to get some canonical constraints, we have two possibilities: 1) $\sigma_-\to\tau$ and $\sigma_+\to\sigma$, and
2) $\sigma_+\to\tau$ and $\sigma_-\to-\sigma$, where $\sigma_\pm=\frac{1}{2}(\tau\pm\sigma)$.

In the first case, $\sigma_-\to\tau$ and $\sigma_+\to\sigma$, the
world-sheet action gets the form
\begin{eqnarray}
S=2\kappa \int_\Sigma d^2\xi \left[x'^\mu \Pi_{+\mu\nu}\dot
x^\nu+2(\bar\theta'^\alpha+\bar\Psi^\alpha_\mu
x'^\mu)(F^{-1})_{\alpha\beta}(\dot\theta^\beta+\Psi^\beta_\nu \dot
x^\nu)\right]\, .
\end{eqnarray}
The canonical momenta conjugated to the variables $x^\mu$, $\theta^\alpha$ and $\bar\theta^\alpha$
\begin{equation}
\pi_\mu=\frac{\partial \mathcal L}{\partial \dot x^\mu}=2\kappa\left[-\Pi_{-\mu\nu}x'^\nu+\frac{1}{2\kappa}\bar\eta'_\alpha\Psi^\alpha_\mu\right]\, ,
\end{equation}
\begin{equation}
\pi_\alpha=\frac{\partial_L \mathcal L}{\partial \dot\theta^\alpha}=-\bar\eta'_\alpha\, ,\quad \bar\pi_\alpha=\frac{\partial_L \mathcal L}{\partial \dot{\bar\theta}^\alpha}=0\, ,
\end{equation}
do not depend on the $\tau$-derivatives, and consequently, there are primary constraints
\begin{equation}\label{eq:Jm}
J_{-\mu}=j_{-\mu}-\bar\eta'_\alpha\Psi^\alpha_\mu\,
,\quad J_{-\alpha}=\pi_\alpha+\bar\eta'_\alpha\, ,\quad \bar
J_{-\alpha}=\bar\pi_\alpha\, ,
\end{equation}
where we introduce
\begin{equation}
j_{\pm \mu}=\pi_\mu+2\kappa\Pi_{\pm\mu\nu}x'^\nu\, .
\end{equation}

If we use the notation
$J_{-A}=(J_{-\mu},J_{-\alpha},\bar J_{-\alpha})$ and the basic Poisson algebra
\begin{equation}
\left\lbrace x^\mu(\sigma),\pi_\nu(\bar\sigma)\right\rbrace=\delta^\mu{}_\nu\delta(\sigma-\bar\sigma)\, ,\quad  \left\lbrace \theta^\alpha(\sigma),\pi_\beta(\bar\sigma)\right\rbrace=\left\lbrace \bar\theta^\alpha(\sigma),\bar\pi_\beta(\bar\sigma)\right\rbrace=-\delta^\alpha{}_\beta\delta(\sigma-\bar\sigma),
\end{equation}
algebra of the constraints gets
the form
\begin{equation}
\{J_{-A},J_{-B}\}=-2\kappa {}^\star G_{AB}\delta'\, ,
\end{equation}
where
\begin{equation}\label{eq:GAB}
{}^\star G_{AB}=\left(
\begin{array}{ccc}
{}^\star G_{\mu\nu} & \frac{1}{2}{}^\star\bar\Psi_{\mu\gamma} &
\frac{1}{2}({}^\star\Psi^T)_{\mu\delta}\\
\frac{1}{2}({}^\star\bar\Psi^T)_{\alpha\nu} & 0 & \frac{1}{8}({}^\star
F^T)_{\alpha\delta}\\ \frac{1}{2}{}^\star\Psi_{\beta\nu} &
-\frac{1}{8}{}^\star F_{\beta\gamma} & 0
\end{array}\right)\, .
\end{equation}
Let us note that ${}^\star G_{AB}$ obeys graded symmetrization
rule ${}^\star G_{AB}=(-)^{AB}{}^\star G_{BA}$.

In the second case, $\sigma_+\to\tau$ and $\sigma_-\to-\sigma$, the
action is of the form
\begin{eqnarray}
S=-2\kappa \int_\Sigma d^2\xi \left[\dot x^\mu \Pi_{+\mu\nu}
x'^\nu+2(\dot{\bar\theta}^\alpha+\bar\Psi^\alpha_\mu \dot
x^\mu)(F^{-1})_{\alpha\beta}(\theta'^\beta+\Psi^\beta_\nu
x'^\nu)\right]\, .
\end{eqnarray}
Similarly, we obtain primary constraints
\begin{equation}\label{eq:Jp}
J_{+\mu}=j_{+\mu}+\bar\Psi^\alpha_\mu\eta'_\alpha\,
,\quad J_{+\alpha}=\pi_\alpha\, ,\quad \bar J_{+\alpha}=\bar\pi_\alpha+\eta'_\alpha\, ,
\end{equation}
where the corresponding algebra of the constraints $J_{+A}=(J_{+\mu},J_{+\alpha},\bar J_{+\alpha})$ is
\begin{equation}
\{J_{+A},J_{+B}\}=2\kappa {}^\star G_{AB}\delta'\, ,
\end{equation}
with the same coefficient as in the first case.

It is easy to check that
\begin{equation}
\{J_{+A},J_{-B}\}=0\, ,
\end{equation}
so that we obtain two independent Abelian Kac-Moody algebras
\begin{equation}\label{eq:algebraJ}
\left\lbrace J_{\pm A}, J_{\pm B}\right\rbrace = \pm 2\kappa {}^\star G_{AB}\delta'\, ,\quad \left\lbrace J_{\pm A}, J_{\mp B}\right\rbrace =0\, .
\end{equation}
Note that the algebra of the constraints closes on the fermionic
T-dual background fields (\ref{eq:GBdual})-(\ref{eq:Fdual}) (except ${}^\star B_{\mu\nu}$).

Because the action is linear in time derivative in both cases, the
canonical Hamiltonian density is zero, $\mathcal H_c=0$, and the
total Hamiltonian takes the form
\begin{equation}
H_{T\pm}=\int d\sigma \mathcal H_{T\pm}=\int d\sigma \lambda^A_\pm J_{\pm A}\, ,
\end{equation}
where $\lambda^A$ are Lagrange multipliers. With the help of (\ref{eq:algebraJ}) it is easy to check that
\begin{equation}
\dot J_{\pm A}=\left\lbrace J_{\pm A}, H_{T\pm}\right\rbrace = \mp 2\kappa {}^\star G_{AB}\lambda'^B_{\pm}\, .
\end{equation}
Consequently, there are no more constraints and for $s\det {}^\star G_{AB}\sim \frac{\det {}^\star G}{\det {}^\star F^2}\neq 0$, all constraints, except the zero modes, are of the second class.

\subsection{Canonical structure with time like evolution parameter -- From Kac-Moody to Virasoro algebra}

Following reasons of Ref.\cite{BPS} we are going to
formulate canonical structure with time-like evolution parameter $\tau=\xi^0$ using the structure with light-like ones $\tau=\sigma_+$ and
$\tau=\sigma_-$. We construct energy-momentum tensor components in Sugawara form as bilinear combination of the currents $J_{\pm A}$
\begin{equation}
T_{\pm}=\mp\frac{1}{4\kappa}J_{\pm A}({}^\star G^{-1})^{AB}J_{\pm B}\, ,
\end{equation}
where
\begin{equation}\label{eq:inverzG}
({}^\star G^{-1})^{AB}=\left(\begin{array}{ccc}
G^{\mu\nu} & -\Psi^{\mu\gamma} & -\bar\Psi^{\mu\delta}\\
\Psi^{\alpha\nu} & -\Psi^\alpha_\rho\Psi^{\rho\gamma} & -\frac{1}{2}(F^{\alpha\delta}+2\Psi^\alpha_\rho\bar\Psi^{\rho\delta})\\
\bar\Psi^{\beta\nu} & \frac{1}{2}\left((F^T)^{\beta\gamma}-2\bar\Psi^\beta_\rho \Psi^{\rho\gamma}\right) & -\bar\Psi^\beta_\rho \bar\Psi^{\rho\delta}
\end{array}\right)\, ,
\end{equation}
is inverse of supermatrix ${}^\star G_{AB}$. Note that the
currents $J_{\pm A}$, which was canonical constraints for
lightlike evolution parameter, are not canonical constraints for
timelike evolution parameter. Here, the canonical constraints are
only energy-momentum tensor components. They satisfy two
independent Virasoro algebras
\begin{equation}
\left\lbrace T_{\pm}(\sigma),T_{\pm}(\bar\sigma)\right\rbrace=-\left[ T_{\pm}(\sigma)+T_{\pm}(\bar\sigma)\right]\delta'\, ,\quad \left\lbrace T_{\pm}(\sigma),T_{\mp}(\bar\sigma)\right\rbrace=0\, ,
\end{equation}
which are equivalent to the algebra of world-sheet diffeomorphisms. The Hamiltonian for $\tau=\xi^0$ is given by
\begin{equation}\label{eq:noviH}
H_c=\int d\sigma \mathcal H_c\, ,\quad \mathcal H_c=T_--T_+\, .
\end{equation}

By straightforward calculation we can prove
\begin{equation}\label{eq:HJ}
\left\lbrace H_c,J_{\pm A}\right\rbrace=\mp J'_{\pm A}\, .
\end{equation}

Using (\ref{eq:inverzG}) we can obtain the expressions of
energy-momentum tensors in terms of the components
\begin{eqnarray}\label{eq:noviHam}
&{}& T_{\pm}=\mp\frac{1}{4\kappa}G^{\mu\nu}J_{\pm \mu}J_{\pm \nu}\mp\frac{1}{2\kappa}J_{\pm\alpha}\Psi^{\alpha\mu}J_{\pm\mu}\mp\frac{1}{2\kappa}\bar J_{\pm\alpha}\bar\Psi^{\alpha\mu}J_{\pm\mu} \\ &{}& \pm\frac{1}{4\kappa}J_{\pm\alpha}\Psi^\alpha\Psi^\beta J_{\pm\beta}\pm \frac{1}{4\kappa}J_{\pm\alpha}\left(F^{\alpha\beta}+2\Psi^\alpha\bar\Psi^\beta\right)\bar J_{\pm \beta}\pm \frac{1}{4\kappa}\bar J_{\pm\alpha}\bar\Psi^\alpha\bar\Psi^\beta \bar J_{\pm\beta}\, .\nonumber
\end{eqnarray}
We can check that our construction is equivalent to that of
Refs.\cite{BNBSPLB, bnbsnpb} obtained by prime calculation. Here
we used the relation between currents $J_{\pm A}$ and the current
$I_{\pm\mu}$ introduced in Refs.\cite{BNBSPLB, bnbsnpb}
\begin{equation}
I_{\pm\mu}=J_{\pm\mu}+J_{\pm\alpha}\Psi^\alpha_\mu-\bar\Psi^\alpha_\mu \bar J_{\pm\alpha}\, .
\end{equation}

The currents ${}^\star J^{A}_\pm$, where index is raised by $({}^\star G^{-1})^{AB}$, are of the form
\begin{equation}
{}^\star J^{A}_\pm\equiv ({}^\star G^{-1})^{AB}J_{\pm B}=\left(\begin{array}{c}
{}^\star J^\mu_\pm\\ {}^\star J^\alpha_\pm\\{}^\star \bar J^\beta_\pm
\end{array}\right)
=\left(\begin{array}{c}
J_{\pm}^\mu-\Psi^{\mu\alpha}J_{\pm\alpha}-\bar\Psi^{\mu\alpha}\bar J_{\pm \alpha}\\
\Psi^{\alpha\nu}J_{\pm\nu}-\Psi^\alpha_\mu \Psi^{\beta\mu}J_{\pm\beta}-\frac{1}{2}(F^{\alpha\beta}+2\Psi^{\alpha}_\mu \bar\Psi^{\beta\mu})\bar J_{\pm\beta}\\
\bar\Psi^{\beta\nu}J_{\pm\nu}+\frac{1}{2}(F^{\gamma\beta}-2\bar\Psi^{\beta\mu}\Psi^{\gamma}_\mu)J_{\pm\gamma}-\bar\Psi^{\beta\mu}\bar\Psi^\gamma_\mu \bar J_{\pm \gamma}
\end{array}\right)\, .
\end{equation}
Let us note that ${}^\star J^{\mu}_\pm=G^{\mu\nu}I_{\pm \nu}$.

\section{Boundary conditions as a canonical constraints}
\setcounter{equation}{0}

In this section we will look for such solution of the boundary conditions that corresponding
noncommutativity parameters are just the background fields of the
fermionic T-dual theory (\ref{eq:GBdual})-(\ref{eq:Fdual}).

\subsection{Choice of the boundary conditions and canonical consistency procedure}

Varying the Hamiltonian (\ref{eq:noviH}) we obtain
\begin{equation}\label{eq:korisno2}
\delta H_c=\delta H_c^{(R)}-\left[\tilde\gamma_\mu^{(0)}\delta x^\mu+\frac{1}{4\kappa}J_{+\alpha}F^{\alpha\beta}\delta \eta_\beta+\frac{1}{4\kappa}\delta\bar\eta_\alpha F^{\alpha\beta}\bar J_{-\alpha}\right]|_0^\pi\, ,
\end{equation}
where $\delta H_c^{(R)}$ is regular term, without derivatives of coordinates and momenta variations, and
\begin{equation}
\tilde\gamma_\mu^{(0)}=\Pi_{+\mu\nu}{}^\star J^\nu_-+\Pi_{-\mu\nu}{}^\star J^\nu_+\, .
\end{equation}
Because the Hamiltonian is time translation generator it must have
well defined functional derivatives with respect to the
coordinates and momenta. Consequently, we get the boundary condition
\begin{equation}\label{eq:GU}
\left[\tilde\gamma_\mu^{(0)}\delta x^\mu+\frac{1}{4\kappa}J_{+\alpha}F^{\alpha\beta}\delta \eta_\beta+\frac{1}{4\kappa}\delta\bar\eta_\alpha F^{\alpha\beta}\bar J_{-\alpha}\right]|_0^\pi=0\, .
\end{equation}

We will choose Dirichlet boundary conditions (fixed string endpoints)
\begin{equation}\label{eq:dbc}
x^\mu|_0^\pi=const.\, ,\quad \eta_\alpha|_0^\pi=const.\, ,\quad \bar\eta_\alpha|_0^\pi=const.\, ,
\end{equation}
which solve boundary condition (\ref{eq:GU}).
They can be expressed in more suitable form in terms of the currents
\begin{eqnarray}\label{eq:gu1}
&{}&\gamma_\mu^{(0)}|_0^\pi=0\, ,\quad \gamma_\mu^{(0)}\equiv J_{+\mu}+J_{-\mu}\, ,\nonumber\\
&{}&\gamma_\alpha^{(0)}|_0^\pi=0\, ,\quad \gamma_\alpha^{(0)}\equiv J_{+\alpha}+J_{-\alpha}\, ,\\
&{}&\bar\gamma_\alpha^{(0)}|_0^\pi=0\, ,\quad \bar\gamma_\alpha^{(0)}\equiv \bar J_{+\alpha}+\bar J_{-\alpha}\, .\nonumber
\end{eqnarray}
In fact, on the equations of motion for momenta $\pi_\mu$, $\pi_\alpha$ and $\bar\pi_\alpha$ we have
\begin{equation}
J_{\pm \mu}=\kappa G_{\mu\nu}\partial_\pm
x^\nu+\frac{1}{2}\bar\Psi^\alpha_\mu \partial_\pm
\eta_\alpha+\frac{1}{2} \partial_\pm
\bar\eta_\alpha\Psi^\alpha_\mu\, ,\quad
J_{\pm\alpha}=-\frac{1}{2}\partial_\pm\bar\eta_\alpha\, ,\quad
\bar J_{\pm\alpha}=\frac{1}{2}\partial_{\pm}\eta_\alpha\, ,
\end{equation}
which means that string endpoints velocities are zero
\begin{equation}
2\kappa G_{\mu\nu}\dot x^\nu=\gamma_\mu^{(0)}+\gamma_\alpha^{(0)}\Psi^\alpha_\mu-\bar\Psi^\alpha_\mu \bar\gamma_\alpha^{(0)}\, ,\quad -\dot{\bar\eta}_\alpha=\gamma_\alpha^{(0)}\, ,\quad \dot\eta_\alpha=\bar\gamma_\alpha^{(0)}\, .
\end{equation}

Following method developed in Refs.\cite{kanonski,BNBS} we will consider the expressions $\gamma_A^{(0)}=(\gamma_\mu^{(0)}, \gamma_\alpha^{(0)}, \bar\gamma_\alpha^{(0)})$ as the canonical constraints.  Applying Dirac consistency procedure we obtain infinite set of the constraints
\begin{equation}
\gamma^{(n)}_A|_0^\pi=0\, ,\quad \gamma_A^{(n)}\equiv \left\lbrace H_c, \gamma_A^{(n-1)} \right\rbrace\, , \,\,\, (n=1,2,3,\dots)
\end{equation}
where
\begin{eqnarray}
\gamma_\mu^{(n)}&=&(-1)^{n}\partial_\sigma^{n} J_{+ \mu}+\partial_\sigma^{n}J_{-\mu}\, ,\nonumber \\
\gamma_\alpha^{(n)}&=&(-1)^{n}\partial_\sigma^{n} J_{+ \alpha}+\partial_\sigma^{n}J_{-\alpha}\, ,\quad (n=1,2,3,\dots)\\
\bar\gamma_\alpha^{(n)}&=&(-1)^{n}\partial_\sigma^{n} \bar J_{+ \alpha}+\partial_\sigma^{n}\bar J_{-\alpha}\, .\nonumber
\end{eqnarray}
With the help of the relation (\ref{eq:HJ}), using Taylor expansion
\begin{equation}
\Gamma_A(\sigma)=\sum_{n=0}^\infty \frac{\sigma^n}{n!}\gamma_A^{(n)}|_0\, ,\quad \left[ \Gamma_A=(\Gamma_\mu,\Gamma_\alpha,\bar\Gamma_\alpha)\right]
\end{equation}
we rewrite these infinite sets of consistency conditions at $\sigma=0$ in compact, $\sigma$ dependent form
\begin{equation}\label{eq:Gx}
\Gamma_\mu(\sigma)=J_{+\mu}(-\sigma)+J_{-\mu}(\sigma)\, ,
\end{equation}
\begin{equation}\label{eq:Ga}
\Gamma_\alpha(\sigma)=J_{+\alpha}(-\sigma)+J_{-\alpha}(\sigma)\, ,
\end{equation}
\begin{equation}\label{eq:bGa}
\bar\Gamma_\alpha(\sigma)=\bar J_{+\alpha}(-\sigma)+\bar J_\alpha(\sigma)\, .
\end{equation}
In the similar way we can write the consistency conditions at
$\sigma=\pi$. If we impose $2\pi$ periodicity of the canonical
variables, the solution of the constraints at $\sigma=0$ also
solve the constraints at $\sigma=\pi$.

Because of the relation
\begin{equation}
\left\lbrace H_c, \Gamma_A\right\rbrace=\Gamma'_A\approx 0\, ,
\end{equation}
there are no other constraints in the theory and the consistency
procedure is completed.

Using the algebra of the currents (\ref{eq:algebraJ}) we obtain the algebra
of the constraints
\begin{equation}\label{eq:algebraG}
\left\lbrace \Gamma_A(\sigma),\Gamma_B(\bar\sigma) \right\rbrace=-4\kappa {}^\star G_{AB}\delta'\, ,
\end{equation}
and conclude that they are of the second class because the metric ${}^\star G_{AB}$ defined in (\ref{eq:GAB}) is nonsingular for $\det {}^\star G_{\mu\nu}\neq 0$ and $\det {}^\star F_{\alpha\beta}\neq0$.

In bosonic case the algebra of the constraints closes on bosonic
T-dual fields. In the particular case the algebra of the
constraints (\ref{eq:algebraG}) closes on fermionic T-dual
background fields (\ref{eq:GAB}).

\subsection{Solution of the boundary conditions $\Gamma_A$}

Let us first introduce new variables symmetric and antisymmetric
under world-sheet parity transformation $\Omega:\sigma\to -\sigma$. For bosonic variables and fermionic momenta we use standard
notation \cite{BNBS}
\begin{eqnarray}\label{eq:bv1}
q^\mu(\sigma)&=&P_s x^\mu(\sigma)\, ,\quad \tilde
q^\mu(\sigma)=P_a x^\mu(\sigma)\, ,\\
p_\mu(\sigma)&=&P_s \pi_\mu(\sigma)\, ,\quad
\tilde p_\mu(\sigma)=P_a \pi_\mu(\sigma)\, \\
\tilde p_\alpha(\sigma)&=&P_a \pi_\alpha(\sigma)\, ,\quad
\tilde{\bar p}_\alpha(\sigma)=P_a \bar \pi_\alpha(\sigma)\,  ,\quad
\end{eqnarray}
while for fermionic coordinates we use subscript $a$
\begin{eqnarray}\label{eq:bv2}
\theta_a^\alpha(\sigma)=P_a \theta^\alpha(\sigma)\, ,\quad
\bar \theta_a^\alpha (\sigma)=P_a \bar \theta^\alpha (\sigma)\, ,
\end{eqnarray}
where the projectors on $\Omega$ even and odd parts are
\begin{equation}
P_s=\frac{1}{2}(1+\Omega)\, ,\quad P_a=\frac{1}{2}(1-\Omega)\, .
\end{equation}

Now we are ready to solve the constraint equations
\begin{equation}
\Gamma_\mu(\sigma)=0\, ,\quad \Gamma_\alpha(\sigma)=0\, ,\quad \bar\Gamma_\alpha(\sigma)=0\, .
\end{equation}
We obtain initial variables in terms of the effective ones
\begin{eqnarray}
&{}&x^\mu(\sigma)=\tilde q^\mu(\sigma)\, ,\quad \pi_\mu=\tilde p_\mu-2\kappa\; {}^\star B_{\mu\nu}\tilde q'^\nu+\frac{\kappa}{2}\left({}^\star \bar\Psi_{\alpha\mu}\theta'^\alpha_a+\bar\theta'^\alpha_a\; {}^\star\Psi_{\alpha\mu}\right)\, ,\label{eq:x}\\
&{}&\theta^\alpha(\sigma)=\theta^\alpha_a(\sigma)\, ,\quad \pi_\alpha=\tilde p_\alpha-\frac{\kappa}{8} \bar\theta'^\beta_a\; {}^\star F_{\beta\alpha}+\frac{\kappa}{2}\; {}^\star\bar\Psi_{\alpha\mu}\tilde q'^\mu\, ,\label{eq:teta}\\
&{}&\bar\theta^\alpha(\sigma)=\bar\theta^\alpha_a(\sigma)\, ,\quad \bar\pi_\alpha=\tilde{\bar p}_\alpha-\frac{\kappa}{8}\; {}^\star F_{\alpha\beta}\theta'^\beta_a-\frac{\kappa}{2}\; {}^\star\Psi_{\alpha\mu}\tilde q'^\mu\, ,\label{eq:barteta}
\end{eqnarray}
where the fermionic dual background fields (with stars) are
defined in (\ref{eq:GBdual})-(\ref{eq:Fdual}). We can reexpress these solutions in terms of the
initial background fields too
\begin{eqnarray}\label{eq:xp}
&{}&x^\mu(\sigma)=\tilde q^\mu(\sigma)\, ,\quad \pi_\mu=\tilde p_\mu-2\kappa B_{\mu\nu}\tilde q'^\nu-\frac{1}{2}\bar\Psi^\alpha_\mu (\eta'_a)_\alpha+\frac{1}{2}(\bar\eta'_{a})_\alpha \Psi^\alpha_\mu\, ,\nonumber \\
&{}&\theta^\alpha(\sigma)=\theta^\alpha_a(\sigma)\, ,\quad \pi_\alpha=\tilde p_\alpha-\frac{1}{2}(\bar\eta'_{a})_\alpha\, ,\\
&{}&\bar\theta^\alpha(\sigma)=\bar\theta^\alpha_a(\sigma)\, ,\quad \bar\pi_\alpha=\tilde{\bar p}_\alpha-\frac{1}{2}(\eta'_a)_\alpha\, ,\nonumber
\end{eqnarray}
where
\begin{equation}
(\eta_a)_\alpha\equiv4\kappa (F^{-1})_{\alpha\beta}(\theta_a^\beta+\Psi^\beta_\mu \tilde q^\mu)\, ,\quad (\bar\eta_{a})_\alpha\equiv4\kappa (\bar\theta_a^\beta+\bar\Psi^\beta_\mu \tilde q^\mu)(F^{-1})_{\beta\alpha}\, ,
\end{equation}
are $\Omega$ odd projections of the variables (\ref{eq:etas}). Note that, as a
difference of all previous cases, our basic effective variables
$\tilde q^\mu$, $\tilde p_\mu$, $\theta_a^\alpha$, $\tilde
p_\alpha$, $\bar\theta_a^\alpha$ and $\tilde{\bar p}_\alpha$ are
$\Omega$ odd and the solution for momenta is nontrivial.

From basic Poisson bracket
\begin{equation}
\{x^\mu(\sigma), \pi_\nu(\bar\sigma)\}=\delta^\mu{}_\nu
\delta(\sigma-\bar\sigma)\, ,
\end{equation}
we obtain the corresponding one in $\Omega$ odd subspace
\begin{equation}\label{eq:pz0}
\{\tilde q^\mu(\sigma)\, ,\tilde p_\nu(\bar\sigma)\}=2\delta^\mu{}_\nu
\delta_a(\sigma\, ,\bar\sigma)\, ,
\end{equation}
where
\begin{equation}
\delta_a(\sigma,\bar\sigma)=\frac{1}{2}\left[
\delta(\sigma-\bar\sigma)-\delta(\sigma+\bar\sigma)\right]\, ,
\end{equation}
is antisymmetric delta function. The factor $2$ in front of antisymmetric delta function comes from the fact that $\Omega$-odd functions
on the interval $[-\pi,\pi]$, $\tilde q^\mu$ and $\tilde p_\nu$, are restricted on the interval $[0,\pi]$ (see \cite{milutin}).

Similarly, using basic Poisson algebra of fermionic variables
\begin{equation}
\{\theta^\alpha(\sigma),\pi_\beta(\bar\sigma)\}=\{\bar\theta^\alpha(\sigma),\bar\pi_\beta(\bar\sigma)\}=-\delta^\alpha{}_\beta\delta(\sigma-\bar\sigma)\,
,\quad
\end{equation}
we have
\begin{equation}\label{eq:pz2}
\left\lbrace \theta_a^{\alpha}(\sigma)\,
,\tilde p_{\beta}(\bar\sigma)\right\rbrace =-2\delta^{\alpha}{}_{\beta}
\delta_a(\sigma\, ,\bar\sigma)\, ,\quad \left\lbrace \bar\theta_a^{\alpha}(\sigma)\,
,\tilde{\bar p}_{\beta}(\bar\sigma)\right\rbrace =-2\delta^{\alpha}{}_{\beta}
\delta_a(\sigma\, ,\bar\sigma)\, .
\end{equation}
The momenta $\tilde p_\mu$, $\tilde p_{\alpha}$ and $\tilde{\bar p}_{\alpha}$ are
canonically conjugated to the coordinates $\tilde q^\mu$,
$\theta_a^\alpha$ and $\bar\theta^\alpha_a$, respectively, in $\Omega$ odd subspace.

\section{Momenta noncommutativity relations}
\setcounter{equation}{0}

When the Neumann boundary conditions have been used
\cite{bnbsjhep, bnbsnpb},  the solution for the super momenta was
trivial while the solution for the super coordinates depended not
only on the effective coordinates but also on the effective
momenta. This was a source of the coordinate noncommutativity which
corresponded to the bosonic T-duality. In the present case, with
Dirichlet boundary conditions, the solution for the super
coordinates is trivial while the solution for the super momenta
depends not only on the effective momenta  but also on the
effective coordinates. This is a source of momenta
noncommutativity which will correspond to the fermionic
T-duality.

Instead to calculate Dirac brackets in the initial phase space
associated with constraints $\Gamma_A$, we will calculate the
equivalent brackets in the reduced phase space. We will put the subscript D to distinguish
them from Poisson ones of initial phase space. With the help
of the solution (\ref{eq:x})-(\ref{eq:barteta}) we find that all
supercoordinates are commutative, while the D brackets of
momenta have a form
\begin{equation}
\left\lbrace \pi_\mu(\sigma), \pi_\nu(\bar\sigma)\right\rbrace_D = 4\kappa\; {}^\star B_{\mu\nu}\partial_\sigma \delta(\sigma+\bar\sigma)\, ,
\end{equation}
\begin{equation}
\left\lbrace \pi_\mu(\sigma), \pi_\alpha(\bar\sigma) \right\rbrace_D= \kappa\; {}^\star \bar\Psi_{\mu\alpha}\partial_\sigma \delta(\sigma+\bar\sigma)\, ,
\end{equation}
\begin{equation}
\left\lbrace \pi_\mu(\sigma), \bar\pi_\alpha(\bar\sigma) \right\rbrace_D= -\kappa\; {}^\star \Psi_{\alpha\mu}\partial_\sigma \delta(\sigma+\bar\sigma)\, ,
\end{equation}
\begin{equation}
\left\lbrace \pi_\alpha(\sigma), \bar\pi_\beta(\bar\sigma) \right\rbrace_D= -\frac{\kappa}{4}\;{}^\star F_{\beta\alpha}\partial_\sigma \delta(\sigma+\bar\sigma)\, ,
\end{equation}
\begin{equation}
\left\lbrace \pi_\alpha(\sigma), \pi_\beta(\bar\sigma) \right\rbrace_D =\left\lbrace \bar\pi_\alpha(\sigma), \bar\pi_\beta(\bar\sigma)\right\rbrace_D =0\, .
\end{equation}

If we define the variables
\begin{equation}
\Pi_\mu(\sigma)=\int_0^\sigma d\sigma_1 \pi_\mu(\sigma_1)\, ,\quad \Pi_\alpha=\int_0^\sigma d\sigma_1 \pi_\alpha(\sigma_1)\, ,\quad
\bar \Pi_\alpha=\int_0^\sigma d\sigma_1 \bar \pi_\alpha(\sigma_1)\, ,
\end{equation}
the noncommutativity relations turn to the standard form
\begin{equation}
\left\lbrace \Pi_\mu(\sigma), \Pi_\nu(\bar\sigma)\right\rbrace_D = 4\;\kappa {}^\star B_{\mu\nu}\theta(\sigma+\bar\sigma)\, ,
\end{equation}
\begin{equation}
\left\lbrace \Pi_\mu(\sigma), \Pi_\alpha(\bar\sigma) \right\rbrace_D= \kappa\; {}^\star \bar\Psi_{\mu\alpha}\theta(\sigma+\bar\sigma)\, ,
\end{equation}
\begin{equation}
\left\lbrace \Pi_\mu(\sigma), \bar \Pi_\alpha(\bar\sigma) \right\rbrace_D= -\kappa\; {}^\star \Psi_{\alpha\mu}\theta(\sigma+\bar\sigma)\, ,
\end{equation}
\begin{equation}
\left\lbrace \Pi_\alpha(\sigma), \bar \Pi_\beta(\bar\sigma) \right\rbrace_D= -\frac{\kappa}{4}\; {}^\star F_{\beta\alpha}\theta(\sigma+\bar\sigma)\, ,
\end{equation}
\begin{equation}
\left\lbrace \Pi_\alpha(\sigma), \Pi_\beta(\bar\sigma) \right\rbrace_D =\left\lbrace \bar \Pi_\alpha(\sigma), \bar \Pi_\beta(\bar\sigma)\right\rbrace_D =0\, ,
\end{equation}
where
\begin{equation}\label{eq:fdelt}
\theta(x)=\left\{\begin{array}{ll}
0 & \textrm{if $x=0$}\\
1/2 & \textrm{if $0<x<2\pi$}\, .\\
1 & \textrm{if $x=2\pi$} \end{array}\right .
\end{equation}
Separating the mean value of momenta
$$\Pi_A(\sigma)=\Pi^{mv}_A+\mathcal P_A(\sigma)\, ,\quad \Pi_A^{mv}=\frac{1}{\pi}\int_0^\pi d\sigma \Pi_A(\sigma)\, ,$$
we obtain that only integrals of the momenta at the string endpoints are noncommutative
\begin{equation}
\left\lbrace \mathcal P_\mu(\sigma), \mathcal P_\nu(\bar\sigma)\right\rbrace_D =\Theta_{\mu\nu}\Delta(\sigma+\bar\sigma)\, ,
\end{equation}
\begin{equation}
\left\lbrace \mathcal P_\mu(\sigma), \mathcal P_\alpha(\bar\sigma) \right\rbrace_D=\bar\Theta_{\mu\alpha}\Delta(\sigma+\bar\sigma)\, ,
\end{equation}
\begin{equation}
\left\lbrace \mathcal P_\mu(\sigma), \bar \mathcal P_\alpha(\bar\sigma) \right\rbrace_D= \Theta_{\alpha\mu}\Delta(\sigma+\bar\sigma)\, ,
\end{equation}
\begin{equation}
\left\lbrace \mathcal P_\alpha(\sigma), \bar \mathcal P_\beta(\bar\sigma) \right\rbrace_D=\Theta_{\alpha\beta}\Delta(\sigma+\bar\sigma) \, ,
\end{equation}
\begin{equation}
\left\lbrace \mathcal P_\alpha(\sigma), \mathcal P_\beta(\bar\sigma) \right\rbrace_D =\left\lbrace \bar \mathcal P_\alpha(\sigma), \bar \mathcal P_\beta(\bar\sigma)\right\rbrace_D =0\, ,
\end{equation}
where the noncommutativity parameters are defined as
\begin{equation}
\Theta_{\mu\nu}=2\kappa\; {}^\star B_{\mu\nu}\, ,\quad \bar\Theta_{\mu\alpha}=\frac{\kappa}{2} \;{}^\star \bar\Psi_{\mu\alpha}\, ,\quad \Theta_{\alpha\mu}=-\frac{\kappa}{2} \;{}^\star \Psi_{\alpha\mu}\, ,\quad \Theta_{\alpha\beta}=-\frac{\kappa}{8}\;{}^\star F_{\beta\alpha}\, ,
\end{equation}
and
\begin{equation}\label{eq:fDelt}
\Delta(x)=2\theta(x)-1=\left\{\begin{array}{ll}
-1 & \textrm{if $x=0$}\\
0 & \textrm{if $0<x<2\pi$}\, .\\
1 & \textrm{if $x=2\pi$} \end{array}\right .
\end{equation}
Therefore, all background fields of the fermionic T-dual theory (\ref{eq:GBdual})-(\ref{eq:Fdual}), except ${}^\star G_{\mu\nu}$,
appear as noncommutativity parameters for the solution of boundary conditions (\ref{eq:dbc}).

\section{Concluding remarks}
\setcounter{equation}{0}

In the present article we considered the relationship between
fermionic T-duality and noncommutativity in type IIB superstring
theory. We used the pure spinor formulation of the theory keeping
all terms up to the quadratic ones and neglecting ghost terms. Our
goal was to find such solution of the boundary conditions which will produce fermionic
T-dual fields as noncommutativity parameters.

First, we performed fermionic T-duality in the way described in
Refs.\cite{ferdual}. Comparing initial and dualized theory, we
found the expressions for fermionic T-dual background fields.

Varying the canonical Hamiltonian and demanding that it has well
defined functional derivatives with respect to the coordinates and
momenta, we obtain the boundary condition (\ref{eq:GU}). In order to satisfy them, for all supercoordinates we chose
Dirichlet boundary conditions. Treating these conditions as canonical constraints and reexpressing them in
terms of useful introduced currents we were able to examine
consistency of the constraints. For nonsingular dual metric
${}^\star G_{\mu\nu}$ and nonsingular dual R-R field strength
${}^\star F_{\alpha\beta}$ all constraints are of the second
class.

Instead to use Dirac brackets we solved the second class
constraints. We took $\Omega$ odd parts of canonical variables as
independent effective variables, and expressed the $\Omega$ even
ones in terms of them. We found that the solution of
supercoordinates was trivial, because they depended only on its
$\Omega$ odd projections. The solutions for supermomenta depend
both on effective supercoordinates and effective supermomenta. So,
as a difference of the previous investigations
\cite{kanonski,bnbsjhep,bnbsnpb,BNBS,susyNC,BNBSPLB}, here
supercoordinates are commutative while integrals of supermomenta
are noncommutative. Similar as in previous investigations,
noncommutativity appears only at the string endpoints, and not in
the string interior. Noncommutativity parameters at $\sigma=0$ and
$\sigma=\pi$ have opposite signs.

Let us comment relation between fermionic T-dual background fields
defined in (\ref{eq:GBdual})-(\ref{eq:Fdual}) with noncommutative
parameters corresponding to the Dirichlet boundary conditions
(\ref{eq:dbc}). All noncommutativity parameters, up to the some
constant multipliers, are equal to the fermionic T-dual fields.
Because noncommutativity relations close on
$\Delta(\sigma+\bar\sigma)$ which is symmetric under $\sigma
\leftrightarrow\bar\sigma$, the noncommutativity parameter
symmetric in space-time indices is absent. Therefore, only T-dual
metric tensor ${}^\star G_{\mu\nu}$ does not appear as
noncommutativity parameter. As well as in the previous cases, dual
fields appear in the algebra of constraints (\ref{eq:algebraG}).
Because here the algebra closes on $\partial_\sigma
\delta(\sigma-\bar\sigma)$ which is antisymmetric under $\sigma
\leftrightarrow \bar\sigma$, the background field antisymmetric in
space-time indices is absent. So, the D-brackets of
$\sigma$-dependent constraints $\Gamma_A$ close on all T-dual
background fields except ${}^\star B_{\mu\nu}$.

There is analogy of the obtained result with that of Ref.\cite{bnbsnpb}. We present that in the following table.

\begin{table}[ht]
\begin{tabular}{|c||c|}\hline
{\bf Bosonic} T-duality & {\bf Fermionic} T-duality\\ \hline \hline
{\bf Neumann} boundary conditions & {\bf Dirichlet} boundary conditions \\ for {\bf bosonic} coordinates & for {\bf all} coordinates \\ \hline $\Omega$ {\bf even} effective supercoordinates & $\Omega$ {\bf odd} effective supercoordinates
 \\ \hline {\bf Supercoordinates} noncommutativity & {\bf Supermomenta} noncommutativity \\ \hline
\end{tabular}
\caption{Analogy between bosonic T-duality and coordinates noncommutativity with fermionic T-duality and momenta noncommutativity}
\end{table}

\end{document}